\documentclass[aps,prl,twocolumn,superscriptaddress,showpacs]{revtex4}
\usepackage{amssymb}
\usepackage{amsmath}
\usepackage{graphicx}

\begin{document}

\title{Universal Properties of Ferroelectric Domains}
\author{Igor A. Luk'yanchuk}
\affiliation{Laboratory of Condensed Matter Physics, University of Picardie Jules Verne,
Amiens, 80039, France}
\affiliation{L. D. Landau Institute for Theoretical Physics, Moscow, Russia}
\author{Laurent Lahoche}
\affiliation{Roberval Laboratory, University of Technology of Compiegne, France}
\author{Ana\"{\i}s Sen\'{e}}
\affiliation{Laboratory of Condensed Matter Physics, University of Picardie Jules Verne,
Amiens, 80039, France}
\date{\today }

\begin{abstract}
Basing on Ginzburg-Landau approach we generalize the Kittel theory and
derive the interpolation formula for the temperature evolution of a
multi-domain polarization profile $\mathbf{P}(x,z)$. We resolve the
long-standing problem of the near-surface polarization behavior in
ferroelectric domains and demonstrate the polarization vanishing instead of
usually assumed fractal domain branching. We propose an effective scaling
approach to compare the properties of different domain-containing
ferroelectric plates and films.
\end{abstract}

\pacs{77.80.Bh, 77.55.+f, 77.80.Dj}
\maketitle

\vspace{5cm}

\begin{figure}[!b]
a) \ \ \ \ \includegraphics  [width=3.8cm]{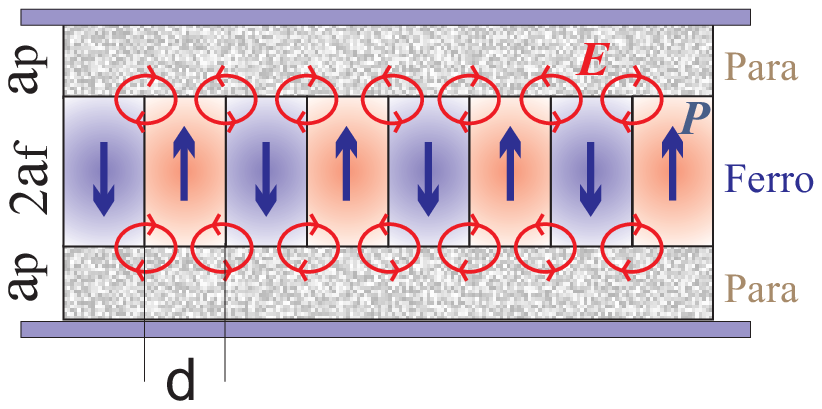} \ \ b) %
\includegraphics  [width=3.2cm]{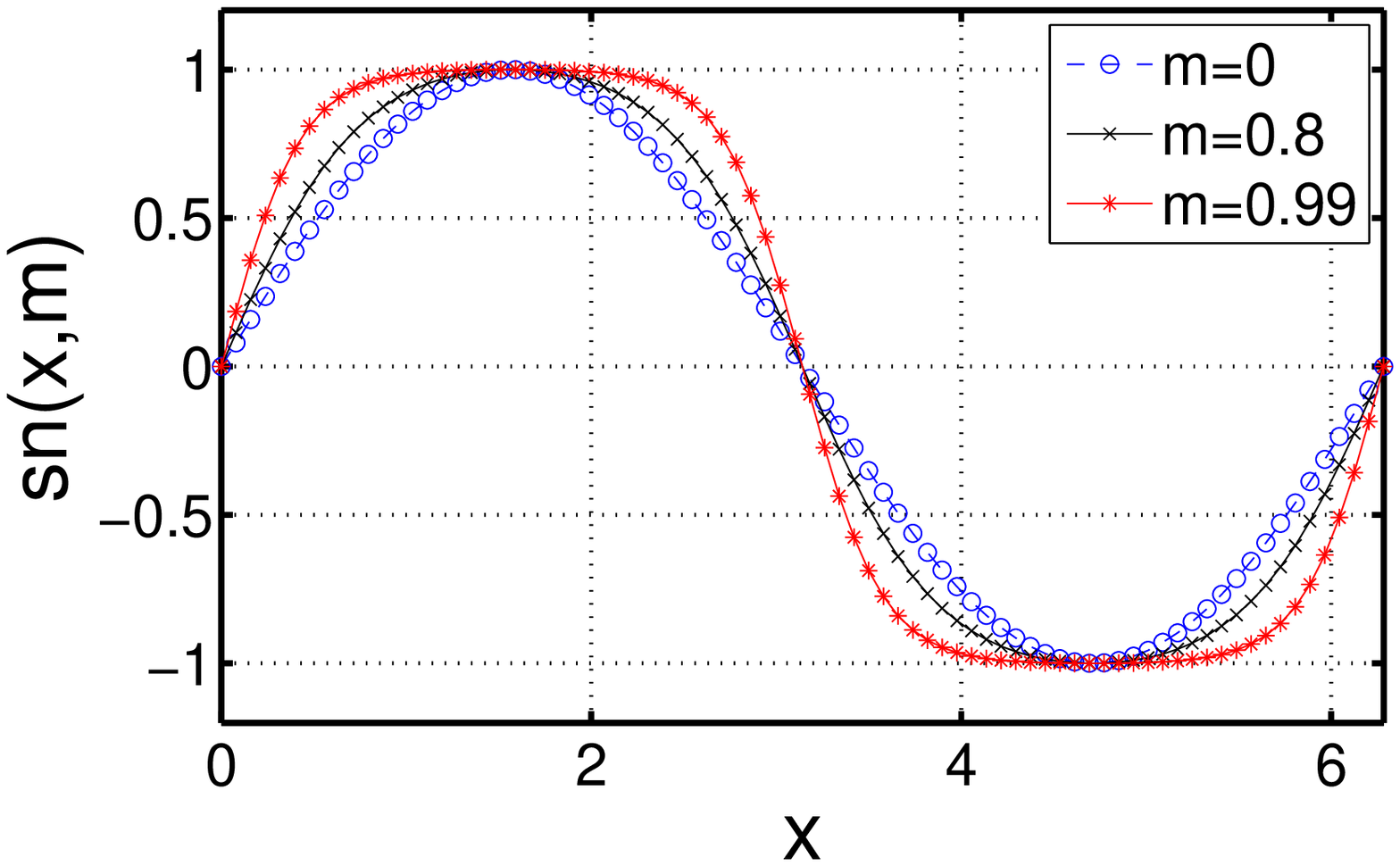} \newline
c) \includegraphics [width=6.3cm] {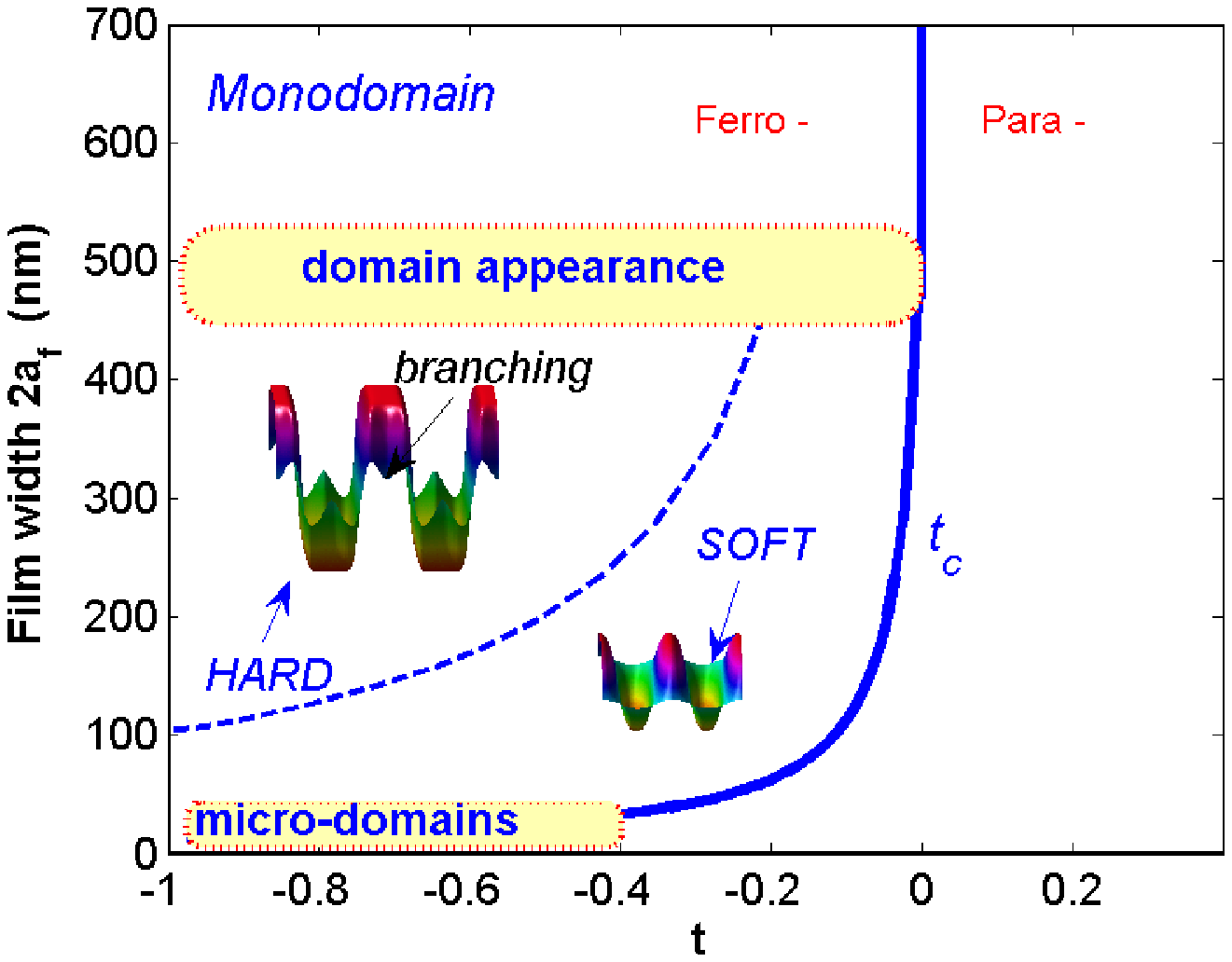} .
\caption{(a)Multi-domain texture of ferroelectric polarization in uniaxial
ferroelectric film, sandwiched by two paraelectric (dead)-layers. The
emerging depolarization electric field is provided by alternating
polarization-induced surface charges and confined in the near-surface layer
of thickness, comparable with domain width $d$. (b) Elliptical functions $y=%
\mathrm{sn}(x,m)$ for different parameters $m$ that we use to model the
domain profile at different $t$. (c) Phase diagram of domain states as
function of sample thickness $2a_f$ and reduced critical temperature $%
t=T/T_{c0}-1$. Polarization profiles of hard and soft domains were obtained
by numerical solution of equations (\protect\ref{Equations})-(\protect\ref%
{perd}). We assume that $\varkappa _{\parallel }\simeq 500$, $\protect%
\varepsilon _{\perp }\simeq 100$, $\protect\varepsilon_p\ll \protect%
\varepsilon _{\perp },\varkappa _{\parallel }$, $\protect\xi_{0x}\simeq 1nm$
and $a_p\simeq30nm$. }
\label{Variat_Fig1}
\end{figure}

Design of ferroelectric devices necessitates taking into account such finite
size effects as the formation of polarization-induced surface charges that,
in turn, produce the energy consuming electrostatic depolarizing fields (see
Ref.\cite{2005_Dawber} for review). As a result, regular periodic structures
of $180^{0}$ domains that alternates the surface charge distribution,
firstly proposed by Landau and Lifshitz\cite{1935_Landau,Landau8} and by
Kittel\cite{1946_Kittel} for ferromagnetic systems, can be formed in
uniaxial easy-axis (natural or stress-induced) ferroelectric plates or films
as an effective mechanism to confine the depolarization field to the
near-surface layer and reduce its energy (Fig.\ref{Variat_Fig1}a). The
energy balance between the field-penetration depth ($\sim $ domain width $d$%
) and domain wall (DW) concentration ($\sim d^{-1}$) leads to the famous
square-root Kittel dependence of $d$ on the film thickness $2a_{f}$ \cite%
{1946_Kittel,2000_Bratkovsky,Guerville,2007Catalan}:
\begin{equation}
d=\sqrt{\gamma \,(\epsilon _{\perp }/\epsilon _{\parallel
})^{1/2}\,(2a_{f}\,\xi _{0x})}\,,\quad \gamma ={\frac{2\sqrt{2}\pi ^{3}}{%
21\zeta (3)}}\simeq 3.53,  \label{Kittel}
\end{equation}%
where $\epsilon _{\parallel }$ and $\epsilon _{\perp }$ are the longitudinal
and transversal dielectric constants and $\xi _{0x}$ is the transverse
coherence length (roughly equal to the DW thickness).

Consider the standard geometry \cite{2000_Bratkovsky} when the uniaxial
ferroelectric film is sandwiched by electroded paraelectric passive layers
of width $a_{p}$ and permittivity $\varepsilon _{p}$. The multi-domain state
should exist in certain intervals of film thickness $2a_{f}$ as shown in
phase diagram in Fig.\ref{Variat_Fig1}c and defined by the condition that
delineates the applicability of Eq.(\ref{Kittel}) and of our further
consideration:
\begin{equation}
\xi _{0x}<d(2a_{f})<a_{p}  \label{condition}
\end{equation}%
the dependence $d(2a_{f})$ being given by (\ref{Kittel}). We also assume the
most realistic case $\varepsilon _{p}\ll \varepsilon _{_{\perp
}}<\varepsilon _{_{\parallel }}$ that gives $d\ll a_{f}$. At this stage the
properties of domain structure do not depend on $a_{p}$, $\varepsilon _{p}$
and electrodes. For thicker films, when $d(2a_{f})$ approaches to $a_{p}$
the emergent depolarizing field interacts with screening electrodes, Eq.(\ref%
{Kittel}) is not valid anymore, $d$ growth exponentially with $a_{p}^{-2}$
\cite{2000_Bratkovsky} and domains practically emerge from the sample.
However in free standing electrodless sample ($a_p\rightarrow \infty$)
Kittel domains can exist in a wider interval of $2a_f$ unless another
restricting mechanism of the internal free charges screening does not came
into the play. For thinner films we are turning to the region of
little-studied atomic-size (microscopic) domains \cite{Bratkovsky2006}.

While domain structures should play a crucial role in the properties of thin
ferroelectric films, only a few theoretical analytical studies of their
temperature dependence have been performed. In particularly the mostly used
Kittel approach \cite{Landau8,1946_Kittel,2000_Bratkovsky} in which the
domain texture is considered as a set of up- and down- oriented (hard)
domains, having a flat polarization profile $\mathbf{P}(x,z)=\pm P_{0}$, DW
are supposed to be infinitely thin and boundary effects on the
ferroelectric-paraelectric interface are neglected, is valid only far below
the transition temperature $T_c$. Although the more general consideration,
proposed by Chensky and Tarasenko (CT)\cite{1982_Chensky} (see also \cite%
{Guerville,2004_Stephanovich}) is based on Ginzburg-Landau equations coupled
with electrostatic equations is valid in the whole temperature interval,
only the solution close to $T_c$ was found.

It is the objective of the present communication to establish the approach
that permits to model the temperature evolution of domain structure. Basing
on CT equations we derive the analytical expression (\ref{varsolut}) for
domain polarization profile that is valid in the whole temperature interval
and includes the Kittel (at low $T$) and CT (at $T=T_c$) solutions as
particular cases. Then, we deduce universal scaling relations between
parameters of the multi-domain state that should be useful in treatment of
experimental data. Our approach is complimentary to the frequently used
first-principia simulations (see e.g. \cite{Bo-Kuai_2006}), that reproduce
the domain structure but give no general vision and parameter dependence of
the results.

Deducing the CT equations we are basing on the Euler-Lagrange variational
formalism, that permits also to obtain the correct boundary conditions as
variation of surface terms. The generating energy functional is written as
\cite{Landau8}:
\begin{equation}
F=\int \widetilde{\Phi }(\mathbf{P},\mathbf{E})dxdz,\quad \widetilde{\Phi }(%
\mathbf{P},\mathbf{E})=\widetilde{\Phi }(\mathbf{P},0)-\mathbf{EP}-\frac{1}{%
8\pi }\mathbf{E}^{2}  \label{Functional}
\end{equation}%
where $\mathbf{E}=(E_{x},E_{z})$, $\mathbf{P}=(P_{x},P_{z})$ and the
field-independent part
\begin{equation}
\widetilde{\Phi }(\mathbf{P},0)=\frac{4\pi }{\varepsilon _{\perp }}\frac{1}{2%
}P_{x}^{2}+\frac{4\pi }{\varepsilon _{i\parallel }}\frac{1}{2}P_{zi}^{2}+%
\frac{4\pi }{\varkappa _{\parallel }}f(P)
\end{equation}%
includes the transversal $P_{x}$, and non-polar longitudinal $P_{zi}$
noncritical contributions ($\varepsilon _{\perp }$,$\varepsilon _{i\parallel
}\gg 1$). The nonlinear Ginzburg-Landau energy depends on the spontaneous $z$%
-oriented polarization $P$ (assuming that $P_{z}=P_{zi}+P$) and is written
as:
\begin{equation}
f(P)=\frac{t}{2}P^{2}+\frac{1}{4}P_{0}^{-2}P^{4}+\frac{\xi _{0x}^{2}}{2}%
\left( \partial _{x}P\right) ^{2}+\frac{\xi _{0z}^{2}}{2}\left( \partial
_{z}P\right) ^{2}  \label{GLD}
\end{equation}%
where the reduced temperature $t$ is expressed via the bulk critical
temperature as: $t=T/T_{c0}-1$, parameter $\varkappa _{\parallel }$ is
expressed via paraelectric Curie constant $C$ and via longitudinal
zero-temperature permittivity $\varepsilon _{\parallel }$ in (\ref{Kittel})
as: $\varkappa _{\parallel }=C/T_{c0}\simeq 2 \varepsilon _{\parallel }$ ,
and coefficient $P_0$ is roughly equal to the saturated bulk polarization at
$T\ll T_c$

The variation of (\ref{Functional}) with respect to $P$ and the
electrostatic potential $\varphi $ ($\mathbf{E}=-\nabla \varphi $) and
excluding of the non-essential variables $P_{x}$ and $P_{zi}$ gives the
system of required equations that describe the ferroelectric transition
taking into account the depolarizing field:
\begin{gather}
(t-\,\xi _{0x}^{2}\partial _{x}^{2}-\xi _{0z}^{2}\partial
_{z}^{2})P+(P/P_{0})^{2}P=-\frac{\varkappa _{\parallel }}{4\pi }\partial
_{z}\varphi ,  \label{Equations} \\
(\varepsilon _{i\parallel }\partial _{z}^{2}+\varepsilon _{\perp }\partial
_{x}^{2})\varphi =4\pi \partial _{z}P.  \notag
\end{gather}%
These equations should be completed by the Poisson equation for paraelectric
media in which ferroelectric film is embedded:
\begin{equation}
(\partial _{z}^{2}+\partial _{x}^{2})\varphi ^{(p)}=0,  \label{fip}
\end{equation}%
and by boundary conditions at the Para-Ferro interface%
\begin{equation}
\varepsilon _{i\parallel }\partial _{z}\varphi -\varepsilon _{p}\partial
_{z}\varphi ^{(p)}=4\pi P,\quad \varphi =\varphi ^{(p)},\quad \partial
_{z}P=0.  \label{bcfi2}
\end{equation}%
that are also obtained as result of variation of (\ref{Functional}) \cite%
{remark}. Periodic conditions
\begin{equation}
P(x,z)=P(x+2d,z)\quad \varphi (x,z)=\varphi (x+2d,z)  \label{perd}
\end{equation}%
with variational parameter $d$ are imposed to describe the periodicity of
domain structure.

A simplification can be achieved if present the initial functional (\ref%
{Functional}) using the dimensionless (prime) variables:
\begin{eqnarray}
z &=&a_{f}\,z^{\prime },\quad x=\tau ^{-1/2}\xi _{0x}\,x^{\prime },\quad
t=\tau \,t^{\prime },  \label{sc1} \\
P &=&\tau ^{1/2}P_{0}\,P^{\prime },\quad \varphi =\frac{1}{\varkappa
_{\parallel }}\,\tau ^{3/2}\,a_{f}P_{0}\,\varphi ^{\prime },  \notag \\
F &=&\frac{a_{f}\,\xi _{0x}}{\varkappa _{\parallel }}\tau
^{3/2}\,P_{0}{}^{2}\,F^{\prime }\,\,\,\,\,\,\,\,\,\,  \notag
\end{eqnarray}%
with
\begin{equation}
\tau =\left( \frac{\varkappa _{\parallel }}{\varepsilon _{\perp }}\right) ^{%
\frac{1}{2}}\frac{\xi _{0x}}{a_{f}}\ll 1
\end{equation}%
in truncated form,
\begin{eqnarray}
F^{\prime } &=&\int [4\pi \left( \frac{1}{2}t^{\prime }P^{\prime 2}+\frac{1}{%
4}P^{\prime 4}+\frac{1}{2}\left( \partial _{x}^{\prime }P^{\prime }\right)
^{2}\right)  \notag  \label{sfn} \\
&&-\frac{1}{8\pi }(\partial _{x}^{\prime }\varphi ^{\prime })^{2}+P^{\prime
}\partial _{z}^{\prime }\varphi ^{\prime }]dx^{\prime }dz^{\prime }
\end{eqnarray}%
that was obtained after neglecting the small terms
\begin{equation}
\widehat{A}_{1}=(\frac{\varepsilon _{\perp }}{\varkappa _{\parallel }})^{1/2}%
\frac{\xi _{0z}}{a_{f}}(\partial _{z}^{\prime }P^{\prime })^{2},\,\,\,\,%
\widehat{A}_{2}=\frac{\varepsilon _{i\parallel }}{\varkappa _{\parallel }}(%
\frac{\varkappa _{\parallel }}{\varepsilon _{\perp }})^{1/2}\frac{\xi _{0x}}{%
a_{f}}(\partial _{z}^{\prime }\varphi ^{\prime })^{2}  \label{small}
\end{equation}%
(justification is given in Appendix) and minimizing over $P_{x},P_{zi}$.

The Euler-Lagrange variation of (\ref{sfn}) over $P^{\prime }$ and $\varphi
^{\prime }$ gives the corresponding dimensionless equations:
\begin{gather}
(t^{\prime }-\,\partial _{x}^{\prime 2})P^{\prime }+P^{\prime 3}=-\frac{1}{%
4\pi }\,\partial _{z}^{\prime }\varphi ^{\prime },  \label{ne1} \\
\partial _{x}^{\prime 2}\varphi ^{\prime }=4\pi \,\partial _{z}^{\prime
}P^{\prime },  \label{ne2}
\end{gather}%
and boundary conditions at $z^\prime_-=0$ and at $z^\prime_+=2a_f^\prime=2$:
\begin{equation}
P^{\prime }=0,\quad \varphi ^{\prime }=\varphi ^{\prime (p)}.  \label{nbc}
\end{equation}
that are simpler then conditions (\ref{bcfi2}) since the order of (\ref%
{Equations}) was reduced by neglecting (\ref{small}). We stress here that
these conditions are \emph{derived} from functional (\ref{sfn}) as
variational surface terms.

Passage to dimensionless variables is the powerful tool that permits to
study the various properties of ferroelectric domains even without solution
the differential equations. Note first that equations (\ref{ne1},\ref{ne2})
contain only one driving variable - the dimensionless temperature $t^{\prime
}$. Therefore the "master" temperature dependence of any physical parameter
calculated from (\ref{ne1},\ref{ne2}) can be re-scaled for any other
ferroelectric sample, using the relations (\ref{sc1}).

We derive now such "master" variational solution of equations (\ref{ne1},\ref%
{ne2}) for domain profile $P^{\prime }(x^{\prime },z^{\prime },t^{\prime })$
valid in the whole temperature interval. Note first that these equations can
be solved analytically close to the transition to a multi-domain
ferroelectric state \cite{1982_Chensky,2004_Stephanovich} that occurs at:
\begin{equation}
t_{c}^{\prime }=-\pi ,\qquad t_{c}=-2\pi \sqrt{\frac{\varkappa _{\parallel }%
}{\varepsilon _{\perp }}}{\frac{\xi _{0x}}{2a_{f}}}  \label{tcc}
\end{equation}%
(in dimensionless and dimensional variables), when polarization has the
sinusoidal (soft) distribution:
\begin{equation}
P^{\prime }(x^{\prime },z^{\prime })=A(t^{\prime })\sin {\frac{\pi x^{\prime
}}{d_{c}^{\prime }}}\sin {{\pi z^{\prime }}}  \label{sinprof}
\end{equation}%
with the half-period $d_{c}^{\prime }=\sqrt{2\pi }$ (that is expressed as (%
\ref{Kittel}) in dimensional variables but with $\gamma =\pi $ and $\epsilon
_{\parallel }=\varkappa _{\parallel }/2$). At lower temperatures domain
walls become sharper due to the admixture of higher harmonics. At lower
temperatures the domains recover the (hard) Kittel-like profile.

To account for both these cases by the unique interpolation formula we shall
exploit the depicted in Fig.~\ref{Variat_Fig1}b periodical elliptical sinus
function $y=\mathrm{sn}(x,m)=\mathrm{sn}(x+4K,m)$, frequently used to
describe the incommensurate phases \cite{Sannikov}. The 1/4 of the
elliptical sinus period is given by the tabled first kind elliptical
integral $K(m)$ \cite{Abramowitz}. The useful property of $\mathrm{sn}(x,m)$
is that, depending on the parameter $0<m<1$ it recovers the all described
above domain regimes: from the soft one (\ref{sinprof}) at $m=0$ when $%
\mathrm{sn}(x,m)\rightarrow \sin x$ (like in Eq. \ref{sinprof}) to the hard
(Kittel-like) one at $m\sim 1$ when $\mathrm{sn}(x,m)\rightarrow $ step-wise
function.

\begin{figure}[t]
\centering a)\includegraphics [width=5cm]{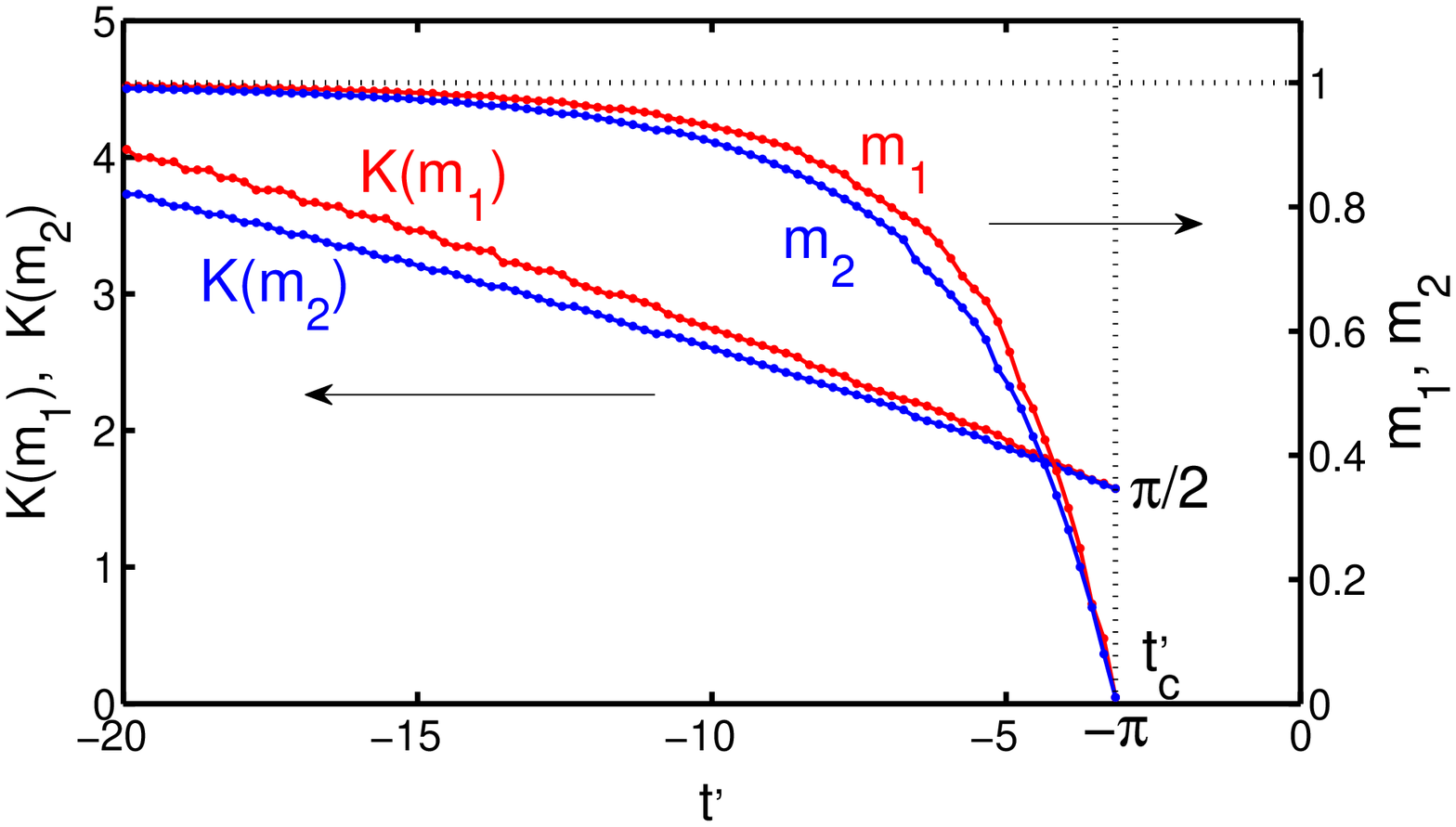} b)%
\includegraphics
[width=5cm]{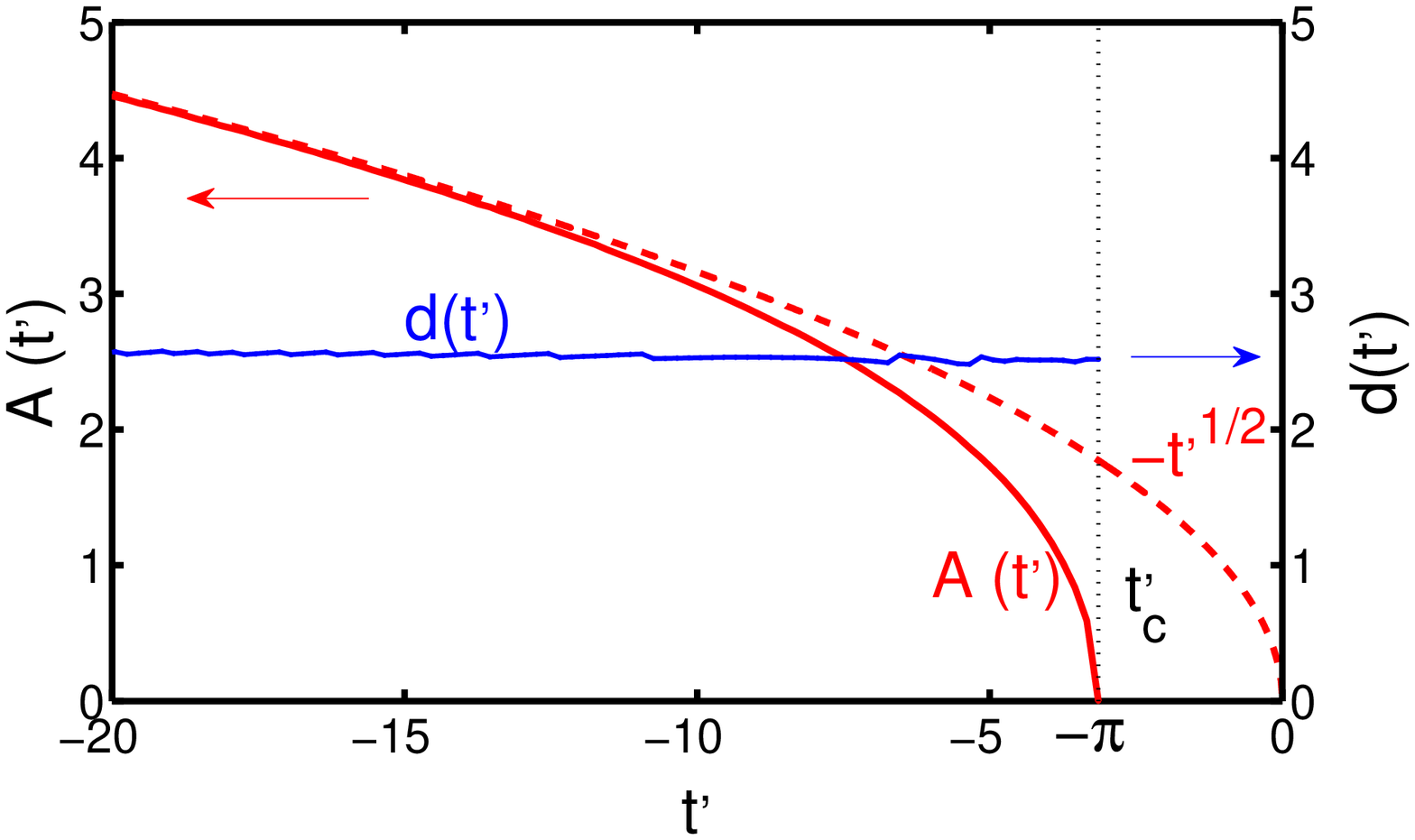}
\caption{Temperature dependencies of parameters of Eq.(\protect\ref{varsolut}%
): (a) elliptic arguments $m_{1}$ and $m_{2}$, elliptic integrals $K_{1}$
and $K_2$ , (b) domain amplitude $A$ and domain lattice period $d^\prime$.
All the variables are dimensionless.}
\label{FigParam}
\end{figure}

After some algebra (justification is given in Appendix) we arrive to the
following variational expression:
\begin{equation}
P^{\prime }=A(t^{\prime })\,\,\mathrm{sn}\left[ \frac{4K_{1}(t^{\prime })}{%
2d^{\prime }(t^{\prime })}\,x^{\prime },\,m_{1}(t^{\prime })\right] \,%
\mathrm{sn}\left[ K_{2}(t^{\prime })\,z^{\prime },m_{2}(t^{\prime })\right]
\label{varsolut}
\end{equation}%
where the temperature dependencies of parameters $m_{1}(t)$ and $m_{2}(t)$,
elliptic integrals $K_{1}(t)$ and $K_{2}(t)$, amplitude $A(t)$ and domain
lattice half-period $d(t)$ are presented in Fig.\ref{FigParam} and for
practical use are approximated as:%
\begin{gather}
A^{\prime }(t^{\prime })\simeq \sqrt{\,t\,\tanh 0.35(t^{\prime
}-t_{c}^{\prime })},\quad d^{\prime }(t^{\prime })\simeq 2.6  \label{approx}
\\
K_{12}(t^{\prime })\simeq 0.85\sqrt{-t^{\prime} },\quad m_{12}(t^{\prime })
\simeq \tanh 0.27(t_{c}^{\prime }-t^{\prime })  \notag
\end{gather}

Formula (\ref{varsolut}) satisfies the boundary conditions
\begin{equation}
P^{\prime }(x^{\prime },z^{\prime })=P^{\prime }(x^{\prime }+2d^{\prime
},z^{\prime }),\quad P^{\prime }(x^{\prime },0)=P^{\prime }(x^{\prime },2)=0,
\label{obc}
\end{equation}%
recovers the soft domain structure (\ref{sinprof}) at $t_{c}^{\prime }$ when
$m_{12}(t_{c}^{\prime })=0$, $A(t^{\prime })\sim (t_{c}^{\prime }-t^{\prime
})^{1/2}$ and the Kittel-like structure at low $t_{c}^{\prime }$ when $%
m_{12}(t_{c}^{\prime })\rightarrow 1$, $A(t^{\prime })\simeq (-t^{\prime
})^{1/2}$, and gives the domain profile at arbitrary $t^{\prime }$.
Parameters $K_{12}(t^{\prime })$ determine the space scale of polarization
variation: in dimensional variables the characteristic domain wall thickness
is $\xi_{x}(t)=\xi_{0x}/(-t)^{1/2}$ whereas the thickness of the
near-surface layer where $P(z)$ restores its equilibrium value is $\sim
d/(-t)^{1/2}\cdot(\varkappa_\parallel/\epsilon_\perp)^{1/2}$ (i.e. $\sim d$
at low $t$).

Variation and vanishing of polarization at the sample surface modifies the
initial assumption of the Kittel model that polarization is permanent inside
domain and resolves the long-standing paradox \cite{Landau8,StrukovLevanyuk}
according to which the permanent domain polarization should be reoriented
close to sample surface by its own depolarization field that exists in the
near-surface layer.

As it follows from our calculations, the nonuniform distribution of
polarization pumps the depolarization charge $\rho(r) \sim \text{div}
\mathbf{P}$ from the sample surface inside the near-surface layer $\sim d$,
reducing the unfavorable depolarization field (justification is given in
Appendix) and its energy $\mathcal{E}_d\sim{E}^2/4\pi\sim4\pi{P}^2$. The
price of this - the dumping of the condensation energy $\mathcal{E}_c\sim
4\pi{P}^2/\epsilon_\parallel$ is not so high because $\epsilon_\parallel \gg
1$. That's why we believe that the near-surface polarization vanishing is
more effective mechanism to overcome the Kittel paradox in ferroelectrics
and reduce the near-surface depolarization energy then the usually assumed
\cite{Landau8,StrukovLevanyuk} but rarely observed fractal branching of
alternatively oriented permanent-polarization domains near the sample
surface.

\begin{figure}[t]
a)\includegraphics [width=2.5cm]{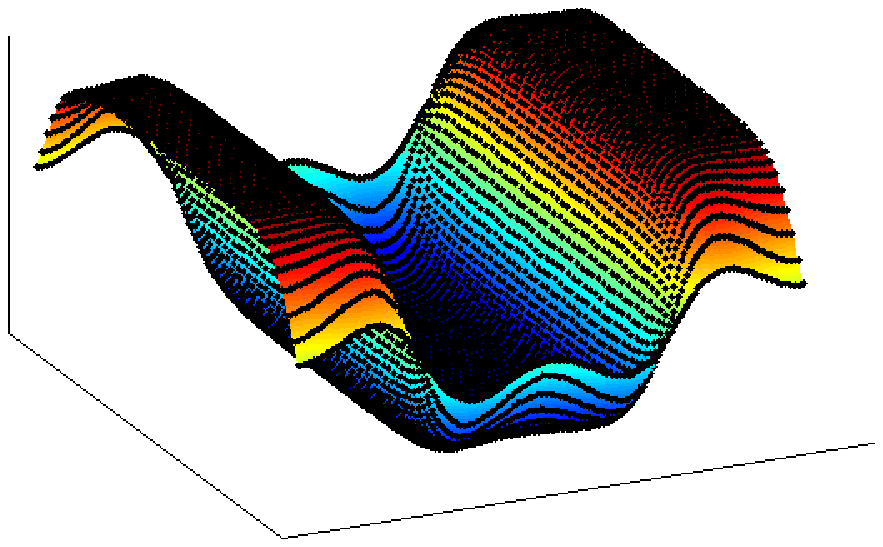} b)%
\includegraphics
[width=2.5cm]{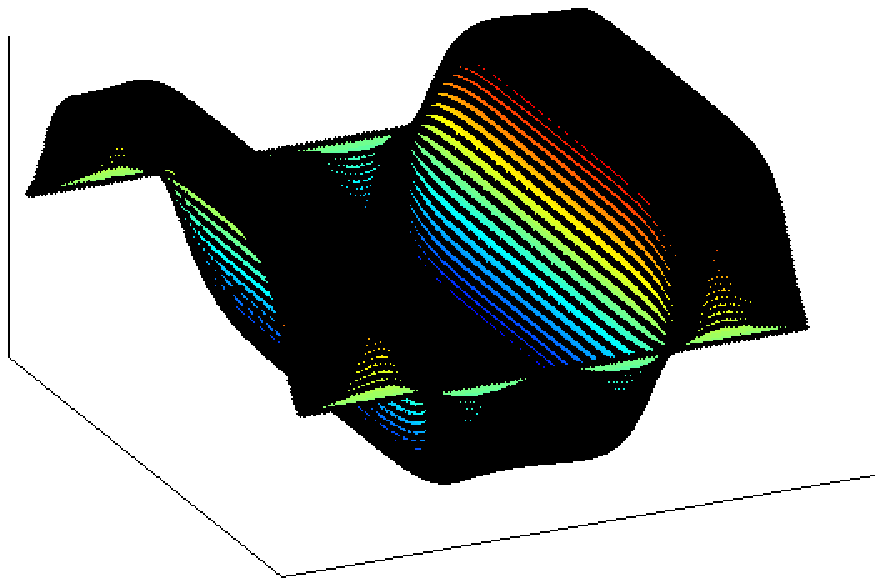}c)\includegraphics
[width=2.5cm]{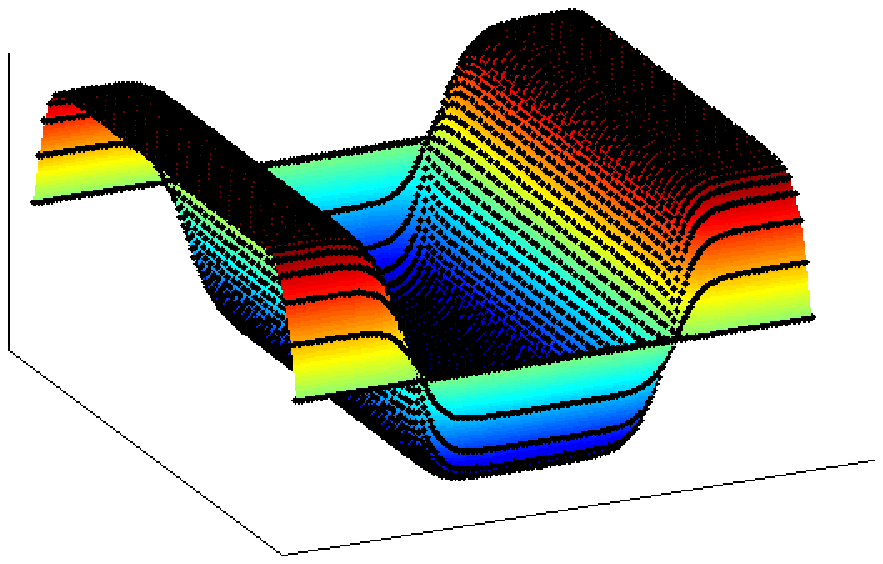}
\caption{Polarization of Kittel domain. (a) Numerical solution of complete
CT equations (\protect\ref{Equations})-(\protect\ref{perd}). (b) Numerical
solution of simplified equations (\protect\ref{ne1})-(\protect\ref{nbc}).
(c) Interpolation formula (\protect\ref{varsolut}) }
\label{FigDomains}
\end{figure}

Polarization decay at the surface is the consequence of the boundary
condition $P^\prime=0$ of simplified equations (\ref{ne1})-(\ref{nbc}). The
validity of this effect is illustrated in Fig.~\ref{FigDomains} where we
compare the numerical solution of simplified equations (\ref{ne1})-(\ref{nbc}%
) (Fig. \ref{FigDomains}b) with that for the complete set of CT equations
(Fig. \ref{FigDomains}a). Clearly the tendency of polarization vanishing is
conserved for the case of general solution in Fig.~\ref{FigDomains}a,
although the "real" boundary condition $\partial _{z}P=0$ (\ref{bcfi2}) is
satisfied exactly at the surface. Interesting to note that the precursor of
the competitive surface domain branching is also seen at Fig.~\ref%
{FigDomains}ab as ripples at the domain end-points. The corresponding
variational solution (\ref{varsolut}) at Fig.~\ref{FigDomains}c is more
smooth, but correctly represents the properties of numerical profile.

We present now several remarkable conclusions about the physical properties
of the multi-domain state which can be obtained only from the scaling
properties (\ref{sc1}), without solution of CT equations (\ref{Equations})-(%
\ref{perd}).

(i) Any transverse length parameter scales as $\tau ^{-1/2}\xi _{0x}$. This,
in particular, justifies the Kittel formula (\ref{Kittel}) for the domain
width $d$ even beyond the flat domain approximation. A convincing
demonstration of the validity of this scaling law was reported recently for
various ferroelectric and ferromagnetic materials \cite{2007Catalan}. The
temperature dependence $d(t)$ can be incorporated into (\ref{Kittel}) as
dependence $\gamma=\gamma(t)$. Meanwhile, the results shown in Fig.~\ref%
{FigParam}b as well as finite-element simulations \cite{Guerville} indicate
that the dependence $d(t)$ is very weak and hence one can extend the
parameter $\gamma \simeq 3.53$ from (\ref{Kittel}) to any temperature.This,
in particular, implies the low temperature hysteresis related with motion of
DW.

(ii) The temperature $t$ scales as $\tau $. Thus, to compare the
domain-provided physical properties of different plates or films (even
constructed from different materials) it should be instructive to trace
their temperature dependencies using the re-scaled coordinate $t/\tau $.

(iii) All the domain-related properties and, in particular, the transition
temperature $t_{c}$ (\ref{tcc}) and the soft-to-hard domain crossover
temperature $t^{\ast }\sim 10t_{c}$ scale as $1/2a_{f}$ with plate (film)
width, as illustrated in Fig.~\ref{Variat_Fig1}c. The temperature interval
for the existence of soft-domains $\Delta t=t_{c}-t^{\ast }$ growth
dramatically with decreasing film thickness and one can expect that for thin
films with $2a_{f}<100nm$ only soft domains with a gradual polarization
distribution are possible.

\null

\null

Summarizing we conclude that domains in \emph{any} ferroelectric
sample and at \emph{any} temperature can be easily obtained from
interpolation formulas (\ref{varsolut},\ref{approx}) applying the
scaling relations (\ref{sc1}). This can be especially helpful to
treat the experimental data, involving the local field distribution
of polarization inside domains like ESR or Raman spectroscopy, TEM
domain imagery etc.

We demonstrated that depending on the temperature and sample width domains
can have soft (gradual) or hard (Kittel) profile. In any case polarization
has the tendency to vanish at sample surface.

Basing on universal scaling relations (\ref{sc1}) we have demonstrated how
the physical properties of the different multi-domain films can be compared
and mapped onto each other. We hope that such method will give the power
tool for analysis and systematization of numerous experimental data for thin
ferroelectric films.

This work was supported by the Region of Picardy, France, by STREP
"Multiceral"(NMP3-CT-2006-032616) and by FP7 IRSES program "Robocon". We
thank to Prof. M. G. Karkut for the useful discussions.

\null

\null

\textbf{APPENDIX }(EPAPS document)

\null

We present here the technical derivation of (i) simplified equations and
corresponding boundary conditions from the generating Euler-Lagrange
functional (ii) interpolation formula for domain polarization (iii)
justification of simplification of generating functional.

\null

We use the defined in the article dimensionless variables, omitting the
prime index.

\null

\textit{(i) Derivation of simplified equations and boundary conditions from
the Euler-Lagrange functional}

\null Euler-Lagrange variation of the simplified dimensionless functional(%
\ref{sfn}) that describes ferroelectric phase in a infinite thin plate
(film) located at $0<z<2$: over polarization $P$ and potential of electric
field $\varphi $ gives:
\begin{gather*}
\delta F=\int \left[
\begin{array}{c}
4\pi \left( tP\delta P+P^{3}\delta P+\left( \partial _{x}P\right) (\partial
_{x}\delta P\right) \\
-\frac{1}{4\pi }(\partial _{x}\varphi )(\partial _{x}\delta \varphi )+\delta
P\partial _{z}+\delta P\partial _{z}\varphi +P\partial _{z}\delta \varphi%
\end{array}%
\right] dxdz \\
=%
\begin{array}{c}
4\pi \int \delta P\left( tP+P^{3}-\partial _{x}^{2}P+\frac{1}{4\pi }\partial
_{z}\varphi \right) dxdz \\
-\int \delta \varphi \left( \frac{1}{4\pi }\partial _{x}^{2}\varphi
-\partial _{z}P\right) dxdz+\left[ \int P\delta \varphi dx\right]
_{z=0}^{z=2}%
\end{array}%
=0
\end{gather*}%
Two first (volume) terms provide the corresponding dimensionless equations (%
\ref{ne1})-(\ref{ne2}) whereas the third (surface) term gives the boundary
condition (\ref{nbc})that should be completed by condition of continuity of
potential at $z=0$ and at $z=2$.

\null

\null

\null

\textit{(ii) Derivation of interpolation formula for domain polarization}

\null
Although the nonlinear equations (\ref{ne1})-(\ref{nbc}) can not be solved
exactly we shall look for their $x$-periodic domain solution in the
variational form
\begin{eqnarray}
P &=&f(z)\,\mathrm{sn}\left[ \frac{4K(m_{1})}{2d}\,x,\,m_{1}\right] ,
\label{variat} \\
\qquad f(0) &=&f(2)=0,\text{ \ }P(x,z)=P(x+d,z)  \notag
\end{eqnarray}%
considering $m_{1}$, $d$ and\ the function $f(z)$ as variational parameters
that minimize (\ref{sfn})

Substitution of (\ref{variat}) back into (\ref{sfn}) and integration over
domain period gives:
\begin{eqnarray}
&&\int [-\frac{1}{8\pi }(\partial _{x}\varphi )^{2}+P\partial _{z}\varphi
]dxdz \\
&&\overset{(\ref{ne2})}{=}\int [-\frac{1}{8\pi }(\partial _{x}\varphi )^{2}-%
\frac{1}{4\pi }\varphi \partial _{x}^{2}\varphi ]dxdz  \notag \\
&&\overset{(\ref{ne2})}{=}2\pi \int \left( \partial _{z}f\right)
^{2}\,(2d)^{2}\,\delta (m_{1})dz  \notag
\end{eqnarray}%
and
\begin{eqnarray*}
&&\int 4\pi \left( \frac{1}{2}tP^{2}+\frac{1}{4}P^{4}+\frac{1}{2}\left(
\partial _{x}P\right) ^{2}\right) dxdz \\
&=&4\pi \int \left(
\begin{array}{c}
\frac{1}{2}tf(z)^{2}\,\alpha (m_{1})+\frac{1}{4}f(z)^{4}\,\eta (m_{1}) \\
+\frac{1}{2}f(z)^{2}\,\frac{4K(m_{1})}{\left( 2d\right) ^{2}}\beta (m_{1})%
\end{array}%
\right) dz
\end{eqnarray*}%
Now the functional depends only on variable $z$:
\begin{equation}
F=4\pi \int \left[
\begin{array}{c}
\frac{1}{2}\,\left( \alpha (m_{1})t+\,\frac{4K(m_{1})}{\left( 2d\right) ^{2}}%
\beta (m_{1})\right) f(z)^{2} \\
+\frac{1}{4}\,\eta (m_{1})f(z)^{4}+\frac{1}{2}\delta (m_{1})\,(2d)^{2}\left(
\partial _{z}f\right) ^{2}%
\end{array}%
\right] dz  \label{Fz}
\end{equation}%
where the coefficients are expressed via complete elliptic integrals of the
first and second kind $K(m)$, $E(m)$ as:%
\begin{eqnarray}
\alpha (m) &=&<\mathrm{sn}^{2}(x,m)> \\
&=&\frac{1}{m}\left[ 1-\frac{E(m)}{K(m)}\right]  \notag \\
\eta (m) &=&<\mathrm{sn}^{4}(x,m)> \\
&=&\frac{1}{3m}\left[ 2\left( 1+m\right) \alpha (m)-1\right]  \notag \\
\delta (m) &=&\frac{\left\langle \mathrm{S}^{2}\left( x,m\right)
\right\rangle }{\left[ 4K(m)\right] ^{2}} \\
&=&\frac{8}{m\left[ 4K(m)\right] ^{2}}\sum_{l=1,3,5}^{\infty }\left[ \frac{1%
}{l}\frac{q^{l/2}(m)}{1-q^{l}(m)}\right] ^{2}  \notag \\
\beta (m) &=&4K(m)\left\langle \left( \mathrm{sn}^{\prime }u\right)
^{2}\right\rangle \\
&=&4K(m)\frac{1}{3}\left[ 2-\left( 1+m\right) \alpha (m)\right]  \notag
\end{eqnarray}%
Here $q(m)=e^{-\frac{K(1-m)}{K(m)}\pi }$, $\mathrm{S}\left( x,m\right)
=\int^{x}\mathrm{sn}\left( u,m\right) du$ and $\left\langle \ldots
\right\rangle $ is the average over the period.

\begin{figure}[t]
\centering \includegraphics [width=5cm] {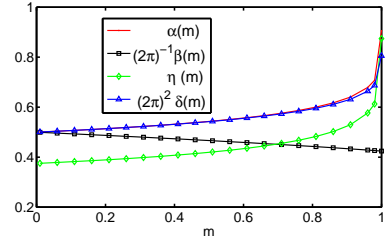}
\caption{ Coefficients $\protect\alpha(m)$, $\protect\beta(m)$, $\protect%
\gamma(m)$ and $\protect\delta(m)$ that enter into the variational
functional (\protect\ref{Fz})}
\label{Variat_Fig4}
\end{figure}

Dependencies $\alpha (m)$, $\beta (m)$, $\eta (m)$ and $\delta (m)$ are
presented in Fig \ref{Variat_Fig4}.

The variational Euler-Lagrange minimum of (\ref{Fz}) is given by the
function:%
\begin{equation}
f(z)=A(t,m_{1},m_{2})\,\mathrm{sn}\left[ K(m_{2})\,z,m_{2}\right]
\label{ff}
\end{equation}%
with
\begin{equation}
A(t,m_{1},m_{2})=2d\left( t,m_{1},m_{2}\right) K(m_{2})\,\sqrt{2\frac{\delta
(m_{1})}{\eta (m_{1})}m_{2}}  \label{Amp}
\end{equation}%
that matches the boundary conditions $f(0)=f(2)=0$ providing that the
dependence $d(t,m_{1},m_{2})$ is fixed by biquadratic equation:
\begin{equation}
\begin{array}{c}
(2d)^{4}\delta (m_{1})(1+m_{2})K^{2}(m_{2})+(2d)^{2}\alpha (m_{1})t \\
+4K(m_{1})\beta (m_{1})%
\end{array}%
=0
\end{equation}%
Substitution of (\ref{ff}) back into (\ref{Fz}) gives:%
\begin{eqnarray}
F(m_{1},m_{2}) &=&-4\pi \frac{1}{4}\eta (m_{1})\int f^{4}(z)\,dz
\label{last} \\
&=&-\frac{1}{2}\pi \eta (m_{1})A^{4}(t,m_{1},m_{2})\eta (m_{2})  \notag
\end{eqnarray}%
Collecting all the results, we present the final variational solution (\ref%
{varsolut}).

\null

\textit{(iii) Justification of simplification of generating functional}

\null

The simplified functional (\ref{sfn}) was obtained by neglecting the terms (%
\ref{small}).

Using profile $P^{ }(x^{},z^{ })$ from (\ref{varsolut}) we can now justify
that contribution of these terms is indeed small by noting that their action
is concentrated in the near-surface layer of thickness $\xi _{r}^{ }\sim
1/K_{2}(t)\sim |t|^{-1/2}$. We will consider only the Kittel regime far from
$t_c=-\pi$. The soft regime close to $t_c$ was already considered in \cite%
{Stephano} .

The relative contribution of the first term to $F$ is estimated as
\begin{equation}
\int \widehat{A}_{1}dx^{{}}dz^{{}}/F\sim (\frac{\varepsilon _{\perp }}{%
\varkappa _{\parallel }})^{1/2}\frac{\xi _{0z}}{a_{f}}\xi _{r}\sim (\frac{d}{%
a_{f}})^{2}|t|^{-1/2}\ll 1
\end{equation}%
that is small for the Kittel domains with $d\ll a_{f}$. Note, however, that
this criteria is not satisfied for monodomain polarization profile that
formally is achieved when $d\rightarrow \infty $. This means that
dimensionless equations (\ref{ne1},\ref{ne2}) can not be applied for
monodomain x-independent solution, that however is unstable towards domain
formation anyway.

Another term $\widehat{A}_{2}$ is related with the energy of the
depolarizing electric field $E_{z}$. According to (\ref{ne2}), this field
can be calculated from the polarization profile (\ref{varsolut}) as:
\begin{equation}
E_{z}(x,z)=-\partial _{z}\varphi =-4\pi \partial _{z}^{2}\int
\int^{xx_{1}}P(x_{2},z)dx_{2}dx_{1}.  \label{Fld}
\end{equation}%
It follows that the depolarization field $E_{z}$ periodically alternates in $%
x-$direction in anti-phase with $P$ and is located in the near-surface layer
of thickness $\xi _{r}$. It vanishes at the surface and in the bulk.
Estimating the maximal value of $E_{z}$ at $x\sim \xi _{r}/2$ as $E_{z\max
}^{{}}\sim A$ we have:
\begin{equation}
\int \widehat{A}_{2}dxdz/F\sim \frac{\varepsilon _{i\parallel }}{\varkappa
_{\parallel }}(\frac{\varkappa _{\parallel }}{\varepsilon _{\perp }})^{1/2}%
\frac{\xi _{0x}}{a_{f}}\xi _{r}\sim \frac{\varepsilon _{i\parallel }}{%
\varkappa _{\parallel }}(\frac{d}{a_{f}})^{2}|t|^{-1/2}\ll 1.
\end{equation}

The physical meaning of this estimation is discussed in the main text of the
article.


\begin{thebibliography}{99}
\bibitem{remark} Frequently used more general boundary condition $\partial
_{z}P=\lambda^{-1} P$ is obtained when polarization is constrained by
additional surface contribution $\sim \lambda^{-1} \int P^2 dx$ to free
energy (\ref{Functional}). We neglect this term here.
\end{thebibliography}
\end{document}